\documentclass{ws-p8-50x6-00}

\begin{document}

\title{Are Bose-Einstein Correlations emerging from correlations of
fluctuations?}  

\author{O.V.Utyuzh and G.Wilk}

\address{The Andrzej So\l tan Institute for Nuclear Studies; 
Ho\.za 69; 00-689 Warsaw, Poland\\E-mail: utyuzh@fuw.edu.pl and
wilk@fuw.edu.pl}

\author{M.Rybczy\'nski and Z.W\L odarczyk}

\address{Institute of Physics, \'Swi\c{e}tokrzyska Academy;
Konopnickiej 15; 25-405 Kielce, Poland\\
E-mail: mryb@pu.kielce.pl and wlod@pu.kielce.pl}


\maketitle

\abstracts{
We demonstrate how Bose-Einstein correlations emerge from the
correlations of fluctuations allowing for their extremely simple and
fast numerical modelling. Both the advantages and limitations of this
new method of implementation of BEC in the numerical modelling of
high energy multiparticle processes are outlined and discussed.
First applications to description of $e^+e^-$ data are given.
} 

\section{Introduction}

Bose-Einstein correlations (BEC) between identical bosons are since
long time recognized as indispensable tool in searching for dynamics
of multiparticle production processes because of their potential
ability to provide some space-time information about them
\footnote{The "infinite" number of references on BEC will be projected
here on some selected number only, starting from the pedagogical
review \cite{ZAJC} and followed by a selected number of more recent,
mostly complementary in their approach to BEC, reviews \cite{BEC}.}.
However, serious investigation of such processes can be performed
only by means of involved numerical modelling using for this purpose
one of the specially designed Monte Carlo event generators (MCEG)
\cite{GEN}. Their {\it a priori} probabilistic structure prevents the
genuine BEC to occur because they are of purely quantum statistical
origin. The best one can do is either to change accordingly the
output of such MCEG's \cite{LS,FW,AFTER,LUND} to provide for the
necessary bunching of identical bosonic particles (mostly $\pi$'s) in
the phase space or to try to incorporate somehow their bosonic
character (or at least some specific features distinguishing 
the bosonic and purely classical particles) into the MCEG itself,
i.e., to construct the MCEG providing on the output particles already
showing the bunching mentioned above \cite{OMT}. Our proposition
discussed here (based on our works \cite{BSWW,NEWBEC}) can be in this
respect regarded as belonging to the first category but, at the
same time, it uses also ideas explored in \cite{OMT} and, to some
extent, can be regarded as improvement of \cite{OMT} (at least in
what concerns the numerical performance). 

Whatever one is doing the final objective is always the same: to
reproduce the characteristic signals of BEC obtained experimentally,
which for the  case of $2$-particle BEC means that the following
two-particle correlation function,
\begin{equation}
C_2(Q=\vert p_i - p_j\vert )\, =\,
             \frac{N_2(p_i,p_j)}{N_1(p_i)\, N_1(p_j)},  \label{eq:C2}
\end{equation}
defined as ratio of the two-particle distributions to the product of
single-particle distributions increases towards $C_2 = 2$ when $Q$
approaches zero. Notice that (\ref{eq:C2}) depends explicitly on the
{\it measured} momenta $(p_i,p_j)$ of the (like) particles, the
searched for space-time information can be obtained only when
treating it as a specific Fourier-transform of the distributions
$\rho(r)$ of the production points, in which case $C_2$ is usually
(schematically) written as\cite{ZAJC,BEC}
\begin{equation}
C_2(Q)\, =\, 1\, +\, \left\vert \int \!dx\, \rho (r)\, e^{iQr}
\right\vert ^2\, =\,  \left\vert \tilde{\rho}(Q)\right\vert ^2 .
\label{eq:C2R} 
\end{equation}
It allows (in principle) to translate the (observed) {\it width} of
the peak in $C_2(Q)$ into the (deduced) {\it size} of the region of
emission $\rho(r)$ \footnote{In reality all this is, of course, much
more complicated \cite{ZAJC,BEC} but this will not be our point of
interest here. Notice only that what is being deduced in this way is
not the whole interaction region but rather the region where the
like-particles with similar momenta are produced. We could call it
the {\it elementary emitting cell}, introducing in this way the idea
used later on in this talk.}. From that point of view the first
approach to the numerical modelling of BEC mentioned above is
addressing (in one or other form) {\it directly} the (\ref{eq:C2R})
whereas second (and our as well) is providing only (\ref{eq:C2})
which can be later analysed (much in the same way as the experimental
data are) to obtain the information on $\rho(r)$ \cite{BEC}.

\section{BEC - quantum-statistical approach}

In what follows we shall treat the BEC as arising because of
correlations of some specific fluctuations present in physical system
under consideration (known as {\it photon bunching} effect in quantum
optics \cite{OPTICS} where similar correlations are also known under
the name of HBT effect). Notice that the main ingredient of $C_2$ is
the correlator $\langle n_1 n_2\rangle$, which can be written as
\cite{ZAJC,F} \footnote{Here $\sigma(n)$ is dispersion of the
multiplicity distribution $P(n)$ of produced secondaries and $\rho$
is the correlation coefficient depending on the type of particles
produced: $\rho = +1,-1,0$ for bosons, fermions and Boltzmann
statistics, respectively.} 
\begin{equation}
\langle n_1 n_2\rangle =  \langle n_1\rangle \langle n_2\rangle
                     + \langle \left(n_1 - \langle n_1\rangle\right)
               \left(n_2 - \langle n_2\rangle\right)\rangle =
                \langle n_1\rangle \langle n_2\rangle
                 + \rho \sigma(n_1)\sigma(n_2). \label{eq:COV}
\end{equation}
Therefore the two-particle correlation function (\ref{eq:C2}) is
entirely given in terms of the covariances (\ref{eq:COV}) explicitly
showing its stochastic character:  
\begin{equation}
C_2(Q=|p_i-p_j|) = \frac{\langle
n_i\left(p_i\right)n_j\left(p_j\right)\rangle}
  {\langle n_i\left(p_i\right)\rangle\langle
n_j\left(p_j\right)\rangle}
       = 1 + \rho \frac{\sigma\left(n_i\right)}
                             {\langle n_i\left(p_i\right)\rangle}\cdot
                        \frac{\sigma\left(n_j\right)}
                        {\langle n_j\left(p_j\right)\rangle} .
\label{eq:algor}
\end{equation}
In eq.(\ref{eq:algor}) above $C_2(Q)$ is just a measure of
correlation of fluctuations present in the system under
consideration, which is maximal (i.e., $C_2 =2$) whenever
fluctuations are maximal (i.e., for $\sigma (n) = \langle n\rangle$,
what happens for the geometrical distribution of the produced
particles, which in turn happens in a natural way where they are
bosons and as such show maximal tendency to bunch themselves in the
same state (or {\it cell}) in the phase space).

This feature is the cornerstone of the only MCEG providing particles
already showing $C_2(Q) > 1$ without any additional procedures
\footnote{The nearest in spirit to is the unpublished attempt
presented in \cite{Cramer}. The two others \cite{Others} are 
providing BEC in not so natural way as \cite{OMT}.}. Those who would
like to start with the symmetrization of the corresponding
multiparticle wave function should realize \cite{ZAL} that such
symmetrization is equivalent to the change from the Maxwell-Boltzmann
statistics (classical, distinguishable particles) to the
Bose-Einstein one (indistinguishable quantum particles). To do this
one has to select as independent subsystems not the individual
particles but rather the groups of particles in the consecutive
states. In effect the original Poissonian distribution of particles
goes over to the geometrical one mentioned above. This can be
visualised in a best way on the following simple example, which will
also represent the main ideas of \cite{OMT} \footnote{Those
interested in details should consult \cite{OMT} directly.}. Suppose
that mass $M$ (in rest) is going to hadronize (for simplicity for one
kind of neutral particles with mass $m$ each). When one is selecting
each time a particle with energy $E_i$ according to the simple
statistical distribution $P(E_i) = \exp \left( -E_i/T\right)$ until
the whole $M$ is used up, one gets multiplicity distribution of the
produced secondaries, $P(n)$, in poissonian form (with $<n>$
depending on the parameter $T$). Suppose now that one is changing
algorithm in a following way: after selection the first energy,
$E_1$, one adds with a probability $P$ to the particle chosen in this
way another particles of {\it the same energy} $E_1$ and does it
until the first failure (i.e., until the random number selected is
greater than $P$). Then one is selecting a new particle with energy
$E_2$ (i.e., in fact one is selecting a new energy $E_2$) and repeats
the above procedure again and again, until the whole mass $M$ is
used up. It is straightforward to realize that what one is getting is
the number of poisson-like distributed {\it cells} (with $\langle
n_{cells}\rangle$ depending on $T$ containing each a number of
geometrically distributed secondaries (with $\langle n\rangle$ in a
given cell given by parameter $P$). Taken together this convolution
of poissonian and geometrical distributions results in well defined 
Negative Binomial (NB) distribution of the produced secondaries
\footnote{This can shed a new light on the NB distributions and its
different applications as discussed during this conference \cite{NB}
with, for example, our {\it cells} being the equivalent to {\it
clans} in the standard approach to NB distributions. Similar concept
of elementary emitting cells has been also proposed in \cite{BSWW}.}.
However, in order to get $C_2(Q)$ with a characteristic width $\Delta
Q \sim 1/R$ (corresponding to a "radius" $R$) one has to allow for a
spreading of energies $E^{(i)}_k$ of particles belonging to the
$k$-th cell, $\Delta_E$. Let us stress at this point that {\it it is
precisely this spreading which translates finally into the
dimensional parameter} $R$ being the main subject of interest when
interpreting experimental results for $C_2$.   

The MCEG proposed in \cite{OMT} has the same structure with particles
selected according to Bose-Einstein distribution, $P(E_i) \sim \exp
\left[ n_i \left(\mu - E_i\right)/T \right]$ ($n_i$ is their
multiplicity and $E_i$ are their energies, the two parameters,
"temperature" $T$ and "chemical potential $\mu$ correspond to the
previous $T$ and $P$) and the like-particles are located in the
cells of a fixed size $\delta y$ (in the longitudinal, i.e., rapidity
$y$ space), which is the third parameter corresponding to the
$\Delta_E$ above). 

It turns out that in this approach one gets at the same time both the
correct BEC pattern (i.e., correlations) and fluctuations (as
characterized by the observed intermittency pattern) \cite{OMT}. This
is very strong advantage of this model, which is so far the only
example of hadronization model, in which Bose-Einstein statistics is
not only included from the very beginning on a single event level,
but it is also properly used in getting the final secondaries. In all
other approaches \cite{LS,FW,AFTER,LUND} at least one of the above
elements is missing. The serious shortcoming of method \cite{OMT}
are, however, numerical difficulties to keep the energy-momentum
conservation as exact as possible and its limitation to the specific
event generator only\footnote{Notice that in our simple example, which
apparently did not show these shortcomings, we were considered only 
one type of produced secondaries. When one attempts to incorporate
all charges and to keep the energy-momentum conserved, our example
starts to be also intractable in practice.}.     

\section{BEC - our proposal: general ideas}

In \cite{NEWBEC} we have proposed a new method of numerical modelling
of BEC, which we shall shortly introduce now (some of its elements
were already formulated in \cite{BSWW}). It was thought as additional
element of any MCEG describing multiparticle production processes. So
far it was tested only on some special hadronization scheme proposed
by us (the CAS model \cite{CAS}) and on the JETSET model for $e^+e^-$
processes (to be shown below). We aimed at fast and universal
algorithm providing BEC among the secondaries produced by a given
MCEG already at the event-by-event basis. 
\vspace{-3mm}
\begin{figure}[h]
\noindent
\begin{center}
\epsfig{file=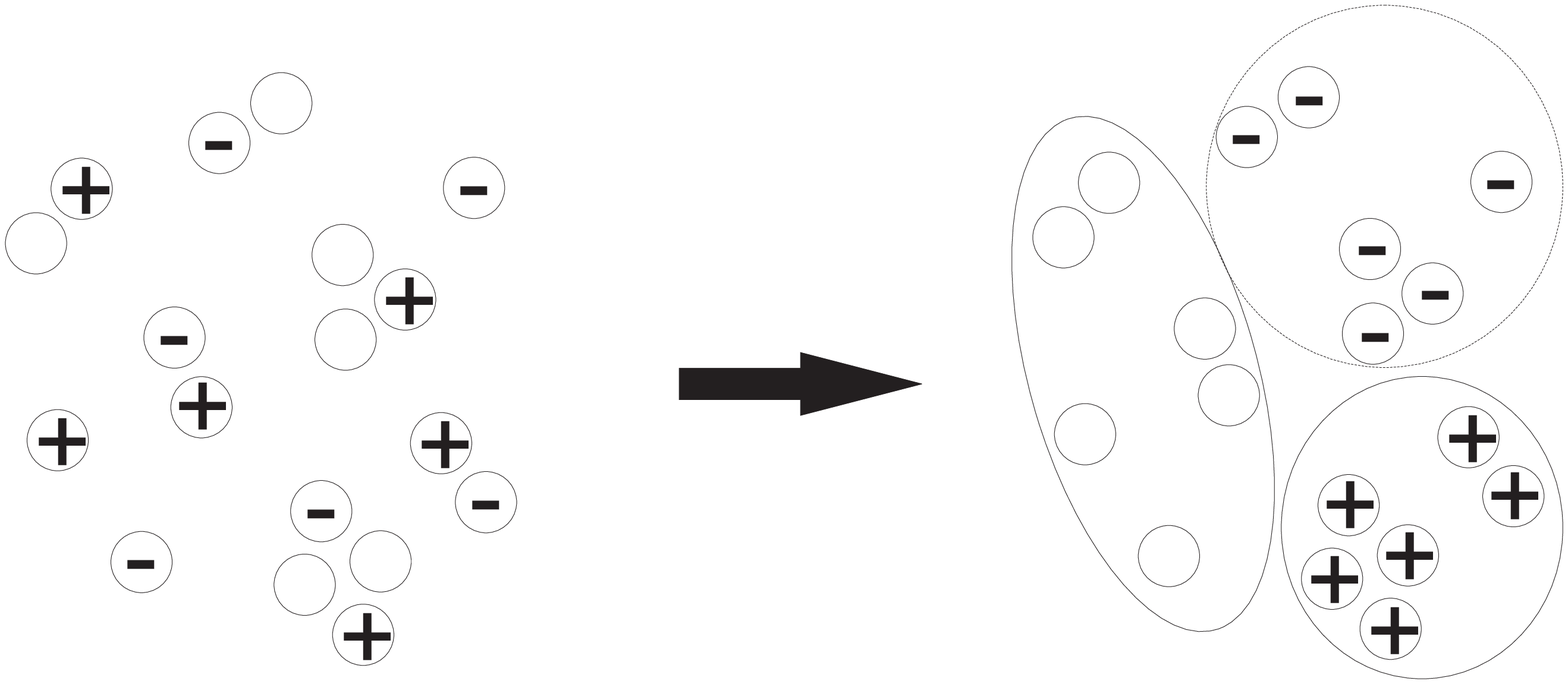, width=60mm}
\end{center}
\vspace{-10mm}
\caption{}
\label{Fig1}
\end{figure}
\vspace{-3mm}

Our way of reasoning was as follows: the BEC is a quantum mechanical
phenomenon whereas all MCEG are of classical character. Therefore to
mimic BEC one has to resign from a part of information provided by
the MCEG, which usually gives us energy-momenta, space-time positions
and charges $(Q_i)$ of the produced secondaries. All methods
\cite{ZAJC,BEC,LS,FW,AFTER,LUND} are in one or other way changing the first
two\footnote{This is especially visible in works claiming to describe
BEC from the very beginning in a quantum mechanical way
\cite{QUANT}.}. We propose to keep them intact and to change,
instead, the original charge assignment provided by the MCEG. This
is done in the following way \cite{NEWBEC}. Suppose that our MC event
generator provides us with $N(+)$, $N(-)$ and $N(0)$ of positive,
negative and neutral particles, uniformly distributed and showing no
BEC pattern (cf. Fig. 1, left panel). We change now their charge
allocation (keeping the same $N(+)$, $N(-)$ and $N(0)$) and getting
the picture shown in the right panel of Fig. 1. The like charges are
in visible (albeit strongly exaggerated) way bunched (correlated)
together leading to signal of BEC. What we have done is the
following: $(a)$ we have resigned from the (not directly measurable)
part of the information provided by event generator concerning the
charge allocation to produced particles  and $(b)$ we have allocated
charges anew in such a way as to keep the like charges as near in
phase space as possible (keeping also the total charge of any kind
the same as the original one). In this way we have formed objects
which we shall call in what follows {\it elementary emitting cells}
(EEC) \cite{BSWW}, each containing only particles of the same sign
and belonging to the same state. In Fig. 1 they are shown as separate
in the phase space but in principle they can overlap. Particles
belonging to such a cell are supposed to be in the same state (the
fact that their momenta differ from each other reflects only the
natural spreading of such state in the momentum space and precisely
this spreading is the source of the analogous complementary spreading
in the position space and eventually in the observed structure of the
$C_2(Q)$).  

The actual implementation of our proposition (algorithm) is presented
in detail in \cite{NEWBEC}. Here we shall discuss some further points
of the proposed approach and present new results of its application
to the known algorithms for hadronization applied to the $e^+e^-$
data. We would like to stress the point that the basic entity here is
not so much the hadronizing source but the EEC mentioned above and
$C_2(Q)$ are in our approach directly sensitive to the number of such
cells and to their mean occupation. There is no direct dependence on
the "size of the hadronizing source" \footnote{Notice, that the
"size" $R$ (or its variants) discussed in all known analyses of BEC
(both data and formulas) are representing at first the distance
between the two emitting points, $R=r_1-r_2$, and therefore are only
indirectly sensitive to the "true" size of the source. Nevertheless,
customary they are referred to as the "size of the source"
\cite{ZAJC}.}. On the other hand our approach contains {\it 
automatically} BEC of {\it all orders} (in practice, the highest
order accounted for is given by the highest multiplicity in a 
given EEC one can reach). 

\section{BEC - our proposal: numerical results}

To illustrate how our prescription works we present in Fig. 2 sample
of results using as initial source of secondaries the CAS
hadronization model developed by us some time ago \cite{CAS} and the
standard JETSET model \cite{JETSET}. This very preliminary and
limited in scope attempt (for example, for simplicity only direct
pions were produced both by the CAS and JETSET MCEG's) shows in a
clear way the main points of our method mentioned before. The EEC's
were formed according to the algorithm given in \cite{NEWBEC} using
as parameter the probability that a new selected particle will joint
the particular EEC which is being actually filled. This probability
was kept constant in the case of CAS (and equal $P$, as indicated in
Fig. 2(a)) whereas in the case of using JETSET it was given by
$P=\exp(-E/T)$ (where $E$ is the energy of the particle considered at
given moment) with value of parameter $T$ as indicated in Fig. 2b. 
In the case of CAS the, so called, "two split-type sources" (cf.
\cite{NEWBEC} for details) were used\footnote{It means that original
source were supposed to consists of two sub-sources of equal mass
each, which were cascading independently, and our algorithm was then
applied to all particles without discriminating which source they
were coming from. In effect one is obtaining much more dense source
(i.e., particles are packed more closely in the phase space) than in
the case of the single cascade only \cite{NEWBEC}. The slope of
$C_2(Q)$ turns out to be very sensitive to this \cite{NEWBEC}. The
opposite situation, in which particles from different sources are
supposed not to show BEC, is referred to as "indep-type sources" and
can, for example, easily explain the BEC puzzle for $W^+W^-$
production data \cite{NEWBEC}.}. Together with fits to the sample of
DELPHI data \cite{DELPHI} are shown the corresponding distributions
of EEC's, $P(N_{cell})$, and particles in such cell, $P(n_{part})$.

\begin{figure}
\begin{center}
\epsfysize=12cm
\centerline{\epsfig{figure=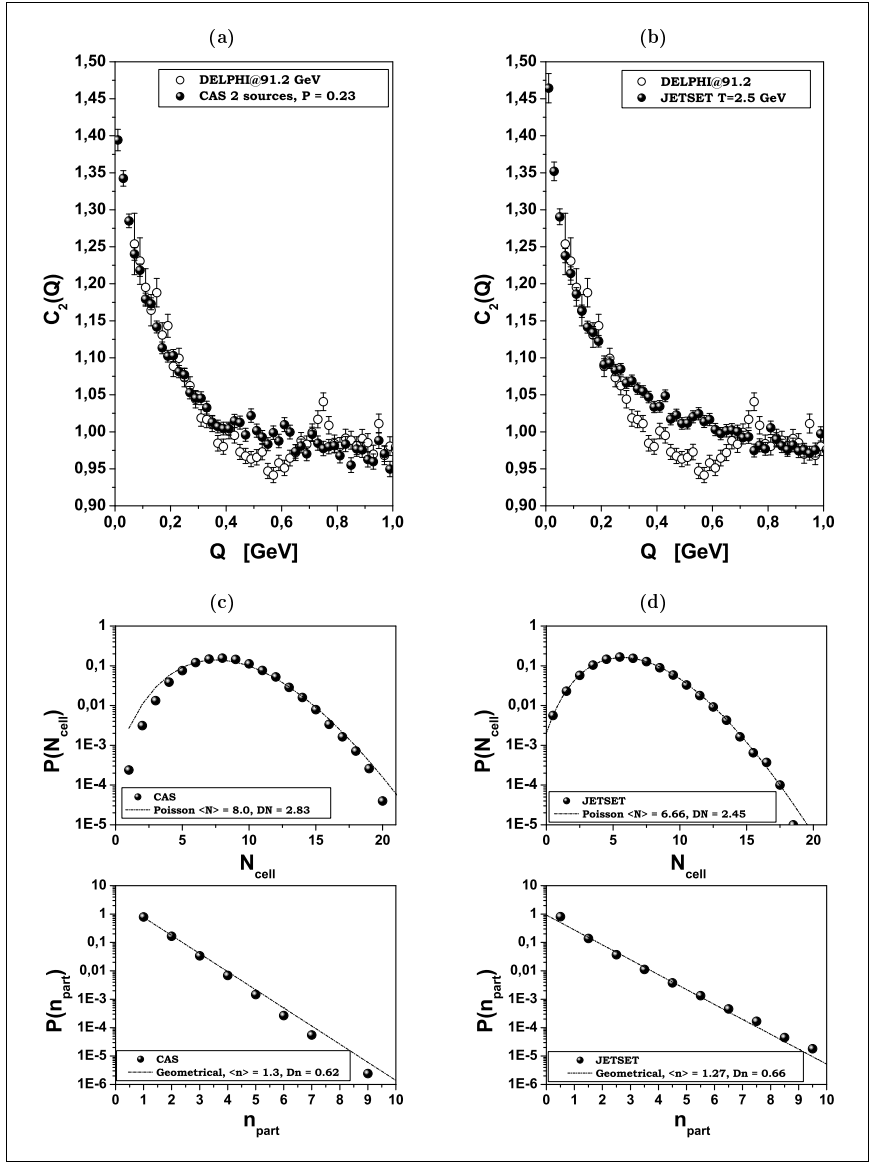,width=11.5cm}}
\caption{Application of our method to fit DELPHI data 
\protect\cite{DELPHI} on $e^+e^-$ annihilation using CAS
\protect\cite{CAS} $(a)$ and JETSET7.4.10 (with standard parameters) 
\protect\cite{JETSET} $(b)$ hadronization models.
 Panels $(c)$ and
$(d)$ show the corresponding distributions of EEC, $P(N_{cell})$, and
distribution of the like-charge particles allocated to such cell,
$P(n_{part})$. The best poissonian and geometrical fits to the,
respectively, $P(N_{cell})$ and $P(n_{part})$ are also shown.}
\end{center}
\end{figure}

Notice that, with a very good accuracy, distribution of particles in
a single EEC is of geometrical type (i.e., according to our
discussion before, corresponding to bosonic particles). This follows
directly from the construction of our algorithm \cite{NEWBEC} (one is
adding new particles to the EEC with probability $P$ until the first
failure, in such a case in the ideal situation $\langle
n_{part}\rangle = P/(1-P)$). The strength of BEC will in our case
depend on how much (on average) particles are allocated to EEC and
how many EEC's one has (on average) in an event. One can see in Fig.
2 that the first number is rather low, $\langle n_{part}\rangle \sim
1.2\div 1.3$, the  probability to get $3$ or more particles in a
single EEC is  of the order of $1\%$ only (and less). It means that
multi-particle BEC are not very prominent feature under normal
circumstances and, if at all, should be probably looked for in a
really very high multiplicity events (which, on the other hand, are
very rare \cite{MAN}). The second number is of the order of $\sim 7$.
Notice that the EEC's are distributed essentially in a poisson-like
manner. It is interesting then to notice that, according to what we
have already mentioned before, it means that the total multiplicity
distribution is of NB type \cite{NB}. Actually, this is the
distribution provided by the original MCEG we are working with (in
our case CAS \cite{CAS} and JETSET \cite{JETSET}), which is then
reproduced by the action of our generator modelling the BEC (as it
should be)\footnote{It is worth to stress at this point that,
according to  our discussion in \cite{BSWW}, a constant number $k$ of
EEC with geometrical distribution of particles in each of them leads
to NB (and in the limit of large $k$ to Poisson) distribution whereas
binomially distributed EEC's, with $k$ limited for some reason, leads
to the so called modified negative binomial (MNB) multiplicity
distributions \cite{MNB} characterised by oscillating cumulant
moments \cite{OSC}.}.  

\section{Summary}

Our presentation of BEC is surely not the orthodox one, i.e., we do
not see BEC as the {\it immediate} source of the information on the
space-time characteristic of the hadronizing object. We approach the
problem from the quantum statistical point of view, stressing more
the behaviour of the like-charged bosons in the momentum-space
(which, after all, is what is really measured!) necessary to be
incorporated in the MCEG whenever one aims to {\it model} the true
BEC. Our particles feel the BEC {\it only} when they are {\it in the
same state} (it allows for the sizeable differences in their momenta
because of the spreading of momenta of the wave packets representing
such states). It means then that our primary objects are EEC's
rather, than the whole emitting hadronic source itself. And it is the
property of the {\it average} EEC, which is responsible for the
finally observed structure of the $C_2(Q)$ function.

Most of our presentation deals with a kind of {\it universal}
algorithm \cite{NEWBEC}, which should be suitable as a simple
addition to (almost) any MCEG \footnote{What are physical changes the
action of our algorithm brings to the original MCEG used is discussed
in full in \cite{NEWBEC}.}. However, it is very likely that such
approach will fail when subjected to the more serious scrutiny than
it is done so far (in the sense that its action will have to be
connected with a re-parameterization of the original MCEG in order to
get the right results, i.e., to correct for deviations introduced by
the BEC implementation \cite{BEC,LS,FW,AFTER,LUND}). In such case, the
only solution left seems to be to start to construct MCEG from
assuring that it models in a correct way the bosonic character of
produced secondaries, i.e., to follow and improve the program started
with the work \cite{OMT} along the lines presented at the beginning
of this talk.

\section*{Acknowledgments}
GW would like to thank Prof. N.G. Antoniou and all Organizers of X-th
International Workshop on Multiparticle Production, Correlations and
Fluctuations in QCD for financial support and kind hospitality.

\end{document}